# Inertia Manipulation through Metric Patching

David Waite

**Abstract**. This paper will present the exact solution for the stress-energy tensor of a spherical matter shell of finite thickness that will patch together different metrics at the boundaries of the shell. The choice of vacuum field solutions for the shell exterior and hollow interior that we make will allow us to manipulate the inertial state of an object within the shell. The choice will cause it to be in a state of acceleration with the shell relative to an external observer for an indefinite time. The stress-energy tensor's solution results in zero ship frame energy requirements, and only finite stress requirements, and we show how any WEC violation can be avoided.

# 1. Introduction

Inertia refers to two distinct concepts. One refers to *inertial mass*. This is the resistance an object has to changes in motion. This is equivalent to the mass m in Newton's second law

$$\mathbf{f} = m\mathbf{a}. \qquad (1)$$

This is then extended to general relativity where it is expressed as an invariant in the relation between two four vectors,

$$F^\lambda = mA^\lambda. \qquad (2)$$

In this context inertia is the resistance to deviation from geodesic motion.

The second concept is the one that this paper is actually concerned with. It is a reference to Newton's first law,

*Not acted on, an object tends to remain in a constant velocity state.*

Extended to general relativity this becomes,

*In the absence of four-forces an object tends to follow geodesic motion.*

Thus in general relativity geodesics exist for which natural motion is not a state of constant velocity. A space-time is here considered for which a ship accelerates away following geodesic motion without any four-forces acting on it. The method we use to construct such a space-time patches together different metrics at matter boundaries.

This method of metric patching may also be useful in investigation into the possibility of other applications where the metric must differ in one region from another.

## 2. Equivalence

Proper acceleration [1] $\alpha$ can be expressed as a function of proper time ct' in terms of $\gamma(\beta)$ and $\beta(ct')$ by the equation

$$\alpha = c^2 \gamma^2 \frac{d\beta}{dct'} \qquad (3)$$

Consider two observers. One will be in an accelerating rocket and the other will be an inertial frame observer. According to the perspective of the ship frame observer, the invariant interval for the space-time we consider is given by

$$ds^2 = \left(1 + \frac{\alpha z'}{c^2}\right)^2 dct'^2 - dx'^2 - dy'^2 - dz'^2 \qquad (4)$$

This results in nonzero affine connections and so according to the rocket frame observer, it is not the rocket that is accelerating. According to the ship frame observer's reckoning, the ship is being held still in the presence of a gravitational force[1] that causes the stars to accelerate in the other direction.

Now we do the following *global* transformation of coordinates for *arbitrarily proper time dependant* acceleration from the accelerated frame to an inertial frame.

$$ct = \int^{ct'} \gamma \, dct' + \gamma \beta z' \qquad (5)$$

$$z = \gamma z' + \int^{ct'} \gamma \beta \, dct' \qquad (6)$$

$$y = y' \qquad (7)$$

---

[1] gravitational forces, inertial forces, forces of affine connection, or fictitious forces are equivalent, given by $f^\lambda = -\Gamma^\lambda{}_{\mu\nu} u^\mu P^\nu$. These are not real forces in the sense that they are not four-vectors and can always be locally transformed away through a transformation to a local free fall frame.

$$x = x' \tag{8}$$

Here γ and β are again expressed as functions of proper time ct'

This globally transforms the invariant interval into the form

$$ds^2 = dct^2 - dx^2 - dy^2 - dz^2 \tag{9}$$

From this inertial frame perspective we see that the rocket is accelerating from among the "stationary" stars and it is accelerating because its exhaust gasses push it.

## 3. The Metric Patching

From here on we will use spherical coordinates. Thus the invariant interval for an accelerated observer can be written

$$ds^2 = \left(1 + \frac{\alpha r \cos\theta}{c^2}\right)^2 dct^2 - dr^2 - r^2\left(d\theta^2 + \sin^2\theta d\phi^2\right) \tag{10}$$

Next lets consider what would happen should we have the case of a piecewise invariant interval as follows.

For r < R

$$ds^2 = dct^2 - dr^2 - r^2\left(d\theta^2 + \sin^2\theta d\phi^2\right) \tag{11}$$

For R < r < R + Δ (The shell's matter region)

$$ds^2 = \left(1 + \frac{\alpha f \cos\theta}{c^2}\right)^2 dct^2 - dr^2 - r^2\left(d\theta^2 + \sin^2\theta d\phi^2\right) \tag{12}$$

For r > R + Δ

$$ds^2 = \left(1 + \frac{\alpha r \cos\theta}{c^2}\right)^2 dct^2 - dr^2 - r^2\left(d\theta^2 + \sin^2\theta d\phi^2\right) \tag{13}$$

A space ship placed in the region r < R no longer needs to keep running its engines to keep from accelerating with the stars because there is no gravitational force in this region.

## 4. The Matter Region's Stress Energy Tensor

Einstein's field equations [2]

$$G^{\mu\nu} = \frac{8\pi G}{c^4} T^{\mu\nu}, \qquad (14)$$

relate the second order differential expressions for the metric in the Einstein tensor $G^{\mu\nu}$ to the energy, momentum, stresses or anything contained in the space given by $T^{\mu\nu}$. Taking the metric for the matter region above, one can either calculate the Einstein tensor by hand and therefor arrive at the required (nonzero) elements of the stress-energy tensor, or one can arrive at the Einstein tensor in a fraction of a second with the aid of a tensor calculus package such as GRTensor II for Maple VI. Either way, the stress energy tensor for the matter region is found to be

$$T^{rr} = \left(\frac{c^4}{8\pi G}\right) \frac{2\alpha \cos\theta \left(r\frac{df}{dr} - f\right)}{c^2 r^2 \sqrt{g_{00}}} \qquad (15)$$

$$T^{r\theta} = T^{\theta r} = \left(\frac{c^4}{8\pi G}\right) \frac{\alpha \sin\theta \left(r\frac{df}{dr} - f\right)}{c^2 r^3 \sqrt{g_{00}}} \qquad (16)$$

$$T^{\theta\theta} = \left(\frac{c^4}{8\pi G}\right) \frac{\alpha \cos\theta \left(r^2 \frac{d^2 f}{dr^2} + r\frac{df}{dr} - f\right)}{c^2 r^4 \sqrt{g_{00}}} \qquad (17)$$

$$T^{\phi\phi} = \left(\frac{c^4}{8\pi G}\right) \frac{\alpha \cos\theta \left(r^2 \frac{d^2 f}{dr^2} + r\frac{df}{dr} - f\right)}{c^2 r^4 \sin^2\theta \sqrt{g_{00}}} \qquad (18)$$

All other $T^{\mu\nu} = 0$

.

The first thing that we note is that $T^{00}$ is zero. Since $T^{00}$ has the interpretation of ship frame energy density we note that there is zero energy requirement for the spaceship to form the desired stress energy tensor of the matter shell. Of course any such shell of matter that can physically be constructed will have mass, but this is a requirement of materials, not of the inertia manipulation scheme.

Second, other types of inertia manipulating spacetimes such as warp drive [3,4] have violated the weak energy condition in that they have required negative energy. This inertia manipulation scheme has no *ship frame* negative energy requirement, but there are frames in which this stress energy tensor, taken as it, is has negative energy density. $T_{\mu\nu}U^{\mu}U^{\nu}$ is less than zero for some time-like vectors. However, there are two ways around the WEC violation issue for this spacetime.

#1 This spacetime was designed to allow for an arbitrarily time dependent proper acceleration and the terms vanish when $\alpha = 0$. Thus, if the acceleration is pulsed at a high rate, any WEC violation can be kept within a short enough time interval so that it is allowed by a quantum mechanics uncertainty principle [5,6,7,8].

#2 Consider a weak field approximation for this spacetime.
$$ds^2 = (1 + 2\phi/c^2)dct^2 - d\sigma^2 \tag{19}$$
As it stands the gravitational potential for the spacetime is $\phi = \alpha f\cos\theta$. The addition of a spherically symmetric mass into the $T^{00}$ term could eliminate any WEC violation altogether, while in the weak field approximation it will only add a term to the potential that is a function of r.
$$\phi_{new} = \alpha f\cos\theta + V(r) \tag{20}$$
This new potential retains the inertia manipulating charachteristics of the old potential merely with the addition of a central attraction.

In construction of such a patched spacetime, one must also be careful about the behavior of the metric at the boundaries. For example, if a function f were chosen so that its first derivative did not match the first derivative of the metric just across the matter boundary then due to the $d^2f/dr^2$ term in the general solution one would wind up with a delta function in the stress-energy tensor. In patching together general spacetimes this can be avoided by requiring the metric and both its first and second derivatives to match across a boundary because this would ensure continuity of the stress energy tensor. It is ensured because Einstein's field equations are second order in the metric. As long as there are no singularities in either the contravariant or covariant metric tensor, one may also avoid divergence in the stress energy tensor at the boundaries in general by matching the metric and its first derivatives across the boundaries. In this case, the stress-energy tensor may not be continuous at the boundaries but it will not diverge.

## 5. Conclusion

Once materials/fields have been found, that can sustain the stresses required, actual experiment in inertia manipulation can be performed. The potential benefits almost go without saying. A ship of this kind maintains its proper acceleration indefinitely, without any energy requirements. Also, if large accelerations are one day found to be achievable the crew of such a ship would not be crushed against the hull of the ship, but would instead accelerate with the ship in a weightless state of free fall.

## 6. Acknowledgements

I would like to thank Prof. R. J. Jacob and Prof. Culbertson for their insights into relativity which helped me in my work leading to this paper